\documentstyle[psfig]{mn}

\newcommand{\rmn}[1]{\mathrm {#1}}

\pagerange{\pageref{firstpage}--\pageref{lastpage}}

\pubyear{1997}

\begin{document}
\label{firstpage}

\title[Hydrodynamics of cooling flows]
{Hydrodynamics and stability of galactic cooling flows}
\author[A. Kritsuk, H. B\"ohringer \& E. M\"uller]{
A. Kritsuk$^{1,2,3}$, H. B\"ohringer$^3$ and E. M\"uller$^2$\\
$^1$ Astronomical Institute, University of St Petersburg, 
Stary Peterhof, St Petersburg 198904, Russia\\
$^2$ Max-Planck-Institut f\"ur Astrophysik, Postfach
1523, D-85740 Garching, Germany\\
$^3$Max-Planck-Institut f\"ur extraterrestrische Physik, 
Postfach 1603, D-85740 Garching, Germany}

\date{Received: February 1998}

\maketitle

\begin{abstract}
Using numerical techniques we studied the global stability of cooling
flows in giant elliptical galaxies.  As an initial equilibrium state
we choose the hydrostatic gas recycling model \cite{kritsuk96}.
Non-equilibrium radiative cooling, stellar mass loss, heating by type
Ia supernovae, distributed mass deposition, and thermal conductivity
are included.  Although the recycling model reproduces the basic X-ray
observables, it appears to be unstable with respect to the development
of inflow or outflow.  In spherically symmetry the inflows are subject
to a central cooling catastrophe, while the outflows saturate in a
form of a subsonic galactic wind.  Two-dimensional axisymmetric random
velocity perturbations of the equilibrium model trigger the onset of a
cooling catastrophe, which develops in an essentially non-spherical
way.  The simulations show a patchy pattern of mass deposition and the
formation of hollow gas jets, which penetrate through the outflow down
to the galaxy core.  The X-ray observables of such a hybrid gas flow
mimic those of the equilibrium recycling model, but the gas
temperature exhibits a central depression. The mass deposition rate
$\dot M$ consists of two contributions of similar size: (i) a
hydrostatic one resembling that of the equilibrium model, and (ii) a
dynamical one which is related to the jets and is more concentrated to
the centre.  For a model galaxy, like NGC~4472, our 2D simulations
predict $\dot M\approx2$~M$_{\odot}$~yr$^{-1}$ within the cooling
radius for the advanced non-linear stage of the instability.  We
discuss the implications of these results to H$\alpha$ nebulae and
star formation in cooling flow galaxies and emphasize the need for
high-resolution 3D simulations.
\end{abstract}


\section{Introduction}
X-ray observations of normal early-type galaxies with the {\em
Einstein} observatory have revealed an extended diffuse thermal
emission from the hot (0.5-2~keV) component of their interstellar
medium [ISM] \cite{forman....79}.  With luminosities
$L_{\rmn{X}}\sim10^{39}$-$10^{42}$~erg~s$^{-1}$ this hot gas
contributes $10^9-10^{10}$~M$_{\odot}$ to the total galaxy mass in its
optical confines (Forman, Jones \& Tucker 1985\nocite{forman..85}).
The X-ray luminosities correlate with the optical (blue) luminosities
of early-type galaxies, although with a large spread of $L_{\rmn{X}}$
at a fixed $L_{\rmn{B}}$ [Canizares, Fabbiano \& Trinchieri
\shortcite{canizares..87}; Donnelly, Faber \& O'Connell
\shortcite{donnelly..90}; Eskridge, Fabbiano \& Kim
\shortcite{eskridge..95b}].  For nearby bright ellipticals the surface
brightness distributions in X-rays and in the optical are similar as
far as the interaction of the galaxy X-ray emitting gas with the
surrounding medium is unimportant (Trinchieri, Fabbiano \& Canizares
1986\nocite{trinchieri..86}).  {\em ROSAT} observations revealed
correlations of the temperature of the hot gas with the stellar
velocity dispersion and with the abundances for a complete sample of
43 elliptical galaxies \cite{davis.96}.  Galaxies with higher velocity
dispersions tend to have higher, approximately solar, abundances and
higher diffuse gas temperatures.  In all cases the gas is
substantially (a factor of $\sim2$) hotter than the kinetic
temperature of the luminous stars indicating the presence of dark
haloes.  The analysis of spectral properties for a sample of 12
early-type galaxies observed with {\em ASCA} has revealed
systematically lower abundances with a mean value of about 0.3 solar
and has confirmed the relationship between stellar velocity dispersion
and gas temperature \cite{matsumoto.....97}.  Gas temperature profiles
for elliptical galaxies determined with {\em ROSAT} and {\em ASCA}
have been found to be surprisingly uniform for projected radii
$r\le10\:r_{\rmn{e}}$ ($r_{\rmn{e}}$ is the so-called effective or
half-light radius).  They exhibit a positive temperature gradient out
to $\sim3\:r_{\rmn{e}}$ followed by a leveling off or gradual decrease
toward larger radii \cite{brighenti.97}.  In addition to hot X-ray
emitting gas early-type galaxies have been shown to have a cooler
component ($10^7-10^8$~M$_{\odot}$) detectable at 100~$\mu$m
\cite{jura...87} as well as warm ionized gas ($10^3-10^5$~M$_{\odot}$
in the central 1-10~kpc) producing optical emission lines
\cite{caldwell84,demoulin-ulrich..84,phillips...86,buson........93,singh...95,macchetto......96}.

Initially, steady-state cooling flow models were invoked to interpret
X-ray observations 
\cite{thomas....86,sarazin.87,sarazin.88,vedder..88,sarazin.89}.  
This, in essence, empirical approach was
based on the simple idea that if the cooling time-scale is shorter
than the Hubble time near the centre, but still longer than the
gravitational time-scale, then a very subsonic quasi-hydrostatic
inflow {\em must} take place.  However, the global stability of these
early models has not been proved yet.  Later time-dependent
hydrodynamic modeling has demonstrated that, in general, such
accretion flows are unstable and suffer cooling catastrophes at their
centres, which were difficult to reconcile with observations
\cite{meiksin88,murray.92}.  Secular variations of the stellar
mass-loss rate and supernova heating rate due to stellar evolution in
the galaxy allowed for a description of the evolution of the 
hot ISM in ellipticals on the Hubble time-scale 
[D'Ercole et al. \shortcite{d'ercole...89};
Ciotti et al. \shortcite{ciotti...91}; David, Forman \& Jones (1990,
1991) \nocite{david..90,david..91a}].  The resulting models evolved
through up to three consecutive evolutionary stages: the wind, outflow
and inflow phases.  Accordingly, most of present-day ellipticals are
in the outflow phase, while a few of the brightest galaxies may have
already experienced their transition to the inflow regime.  Although
the problem of the cooling catastrophe was not resolved, an important
outcome of spherically symmetric evolutionary modeling was the
understanding that the use of steady-state cooling flow models to
diagnose mass accretion rates in dynamical situations can lead to
erroneous results \cite{ciotti...91,murray.92}.

In the absence of direct measurements of the flow velocity a
systematic study of global hydrodynamic and thermal stability
properties of the hot ISM in ellipticals can provide a selection
criterion for realistic flow regimes.  Starting from the essential
physics of cooling flows, we have constructed a hydrostatic
equilibrium model, which describes a stable {\em in situ} recycling of
gas shed by stars through local thermal instabilities into a
condensate, which can become the material for further star formation
\cite{kritsuk96}.

In this paper we probe the global stability of the recycling model in
the inflow-outflow context.  Section 2 describes the basic assumptions
of the hot ISM model and of the underlying galaxy model.  Section 3
gives the details of the numerical method used in the simulations.
The evolution of spherically symmetric perturbations and their
stabilization by a conductive heat flux are considered in Section 4.
In Section 5 we abandon the restrictive assumption of spherical
symmetry and study the development of instabilities by means of
two-dimensional simulations.  This allows for a better insight into
the complex physics of cooling flows.  The discussion and conclusions
can be found in Sections 6 and 7, respectively.

\section{The model}
\subsection{Input physics}
It is usually assumed that the source of the hot ISM in elliptical
galaxies is provided mainly by mass loss in the form of winds from red
giant stars orbiting in the gravitational potential of the stellar
system and its dark halo.  The gas is supposed to be well mixed by
`collisions' of the winds originating from stars, which pass by close
to each other.

Supernova (SN) explosions of type Ia are considered to be a major
potential heating source for the gas, although there is a great
uncertainty in the rate of SNe Ia.  SNe can also play an important
role in the gas enrichment with heavy elements, if the ejecta are well
mixed with the ISM before they cool down considerably.

We assume that evolutionary processes in the stellar system provide a
steady source of small-scale non-linear perturbations, which {\em
unavoidably} become thermally unstable in the temperature regime of
interest.  Heat conduction, which is efficient in such a hot tenuous
plasma, is able to suppress the growth of small-amplitude
perturbations in the short wavelength limit.  However, because of
different non-linear dependencies of the conduction flux and the
radiative cooling rate on local physical conditions in the gas, finite
amplitude small-scale perturbations will still be able to grow
non-linear.  Hence, a fraction of the gas will quasi-isobarically cool
down to at least $T\sim10^4$~K (becoming $\geq10^3$ times denser) and
will drop out from the flow, thereby enhancing the overall cooling of
the gas and removing a fraction of its momentum.

The final fate of the condensing material is still a matter of debate.
If at some later point star formation occurs in the cold condensate,
one can expect that feedback heating by SNe~II and Ib/c will return
part of the energy back to the hot phase (with some delay).  
Currently our simplified model does not incorporate this feedback heating.

Thus, we introduce a set of sources of mass, momentum and energy of
the hot ISM, which are distributed over the whole galaxy.  Their
efficiencies depend both on the local physical conditions in the gas
and on the local parameters of the stellar system.  It is one of the
purposes of our hydrodynamic modeling to follow the evolution of a
hot corona driven by the competition between sources and sinks.  In
several other evolutionary calculations the rates of processes
directly related to stellar evolution (e.g., the SN rate) were assumed
to be variable on a time scale of a few Gyr. However, we keep the
rates constant and concentrate mainly on the short-term evolution of
the cooling flows excluding the early phases of galaxy formation from
our consideration.

Our hydrodynamic description is valid only on scales $l\ge{\cal L} =
\max\{l^{\rmn{wn}}, l^{\rmn{sn}}, l^{\rmn{ti}}\}$, which are larger
than the maximum scale associated with discrete physical sources, such
as a region blown by a stellar wind ($l^{\rmn{wn}}$), a SN remnant
($l^{\rmn{sn}}$), or a cooling condensation ($l^{\rmn{ti}}$).  The
scale ${\cal L}$ depends on the distance from the galactic centre.  We
refer to \cite{mathews90} for a detailed discussion of these issues
[see also \cite{kritsuk92} for details on thermal instabilities].

Below we briefly define the parameters, which control the efficiency
of the sources.

\subsubsection{Stellar mass loss}
The rate of gas supply to the ISM due to stellar mass loss and SNe Ia
is defined as $\alpha \rho_*(r)$, where $\rho_*$ is the stellar mass
density and $\alpha = \alpha_* + \alpha_{\rmn{sn}} \equiv const$.  It
incorporates a contribution from `quiescent' stellar mass loss
$\alpha_* \simeq 4.7 \times 10^{-20}$~s$^{-1}$ 
\cite{faber.76b,sarazin.87} 
with a small addition by supernovae $\alpha_{\rmn{sn}} =
3.1 \times 10^{-21}$~s$^{-1}$ (the estimate corresponds to
$r_{\rmn{sn}} = 0.55\:h_{75}^2$ SNu, $M_{\rmn{sn}} = 1.4\;M_{\odot}$,
and $(M/L_B)_* = 8\;M_{\odot}/L_{\odot}$).  According to van den Bergh
\& Tammann \shortcite{vandenbergh.91} the SNe~Ia rate, $r_{\rmn{sn}}$,
is uncertain by a factor of the order of 1.5, but a significantly
lower estimate for E/S0 galaxies, $r_{\rmn{sn}} = 0.15\pm0.06
\:h_{75}^2$, was derived by Cappellaro et~al.
\shortcite{cappellaro......97}. \footnote{$h_{75}=H_0/75$
km~s$^{-1}$Mpc$^{-1}$.}

\subsubsection{Radiative cooling}
We use a non-equilibrium radiative cooling function $\Lambda(T,
Z_{\odot})$~erg~cm$^3$g$^{-2}$s$^{-1}$ for optically thin hot plasma,
defined in the temperature range $T\in[10^4, 10^{8.5}]$ K, which we
compiled from \cite{sutherland.93} for the case of a solar gas
metallicity $Z_{\odot}$, zero diffuse radiation field, and with
abundance ratios taken from \cite{anders.89}.

\subsubsection{Mass deposition}
We define the rate of mass deposition to be a function of the local
conditions in the hot ISM, $\dot\rho_{\rmn{ti}} = b\chi(n)\rho$, where
$\rho$ is the gas mass density.  The spectrum of primeval small-scale
perturbations in the gas and, in particular, its deformation by heat
conduction are described by a parameter $b \equiv const\in [0,\; 1]$.
The Heaviside function $\chi$ and the linear instability growth rate
$n$ are defined as
\begin{equation}
\chi(n)=\cases{n, &if $n\ge 0$;\cr0, &otherwise;} 
\end{equation}
\begin{equation}
n\equiv {\partial \over \partial t} 
\left(\ln{\delta \rho \over \rho}\right)={1\over
c_p}\left({2\rho\Lambda\over T}-\rho{\rmn{d}\Lambda\over \rmn{d} T}\right) -
{\alpha\rho_*\over \rho}. \label{GrowthRate}
\end{equation}
Here $c_p$ is the specific heat at constant pressure. The first term
in the {\em rhs} of equation (\ref{GrowthRate}) coincides with Field's
instability criterion, the second one describes stabilization due to
stellar mass loss \cite{kritsuk92}.  We assume $b=0.5$ for the models
described in this paper. The motivation for this choice can be found
in \cite{kritsuk96}.

\subsubsection{Distributed heating}
The rate of heating due to thermalization of stellar winds and due to
SNe~Ia is assumed to be equal to $\alpha\rho_*T_0$, where $T_0 =
(\alpha_*T_* + \alpha_{\rmn{sn}}T_{\rmn{sn}})/\alpha$ is the
characteristic temperature of the heat source.  Heating by
thermalization of winds from stars orbiting in the galactic potential
is defined by $T_* = {\mu\sigma_*^2 \over {\cal R}} = 6.47 \times
10^6$~K (we assume $\sigma_*=300$ km/s and a mean molecular weight
$\mu=0.63$; ${\cal R}$ is the gas constant).  The SN temperature
$T_{\rmn{sn}} = {\mu v_{\rmn{ej}}^2 \over3 {\cal R}} = 1.09 \times
10^9$~K for $E_{\rmn{sn}} = 6\times10^{50}$~erg and $M_{\rmn{sn}} =
1.4\; M_{\odot}$.  With these values $T_0 \simeq 7\times10^7$~K, while
for $r_{\rmn{sn}} = 0.15$ SNu the source temperature is lower: $T_0
\simeq 2.6\times10^7$~K. We assume $T_0= 5\times10^7$~K in our
simulations.

\subsubsection{Thermal conduction}
The large-scale heat flux {\boldmath{$q$}}$=-\eta\cdot\kappa\nabla T$
is defined through a reduction factor $\eta\le1$, which is treated as
a free parameter, and the Spitzer \shortcite{spitzer62} conductivity
coefficient for a fully ionized gas 
\begin{equation}
\kappa={1.84\times10^{-5}T_{\rmn{e}}^{5/2}\over \ln{\Lambda_{\rmn{c}}}}\;\;
\rmn{ergs\;s^{-1}\;K^{-1}\;cm^{-1}},
\end{equation}
where for $T > 4.2\times10^5$~K the Coulomb logarithm is
$\ln{\Lambda_{\rmn{c}}} = 32.0+\ln{[n^{-1/2}(T_{\rmn{e}}/10^7}$~K)].

\subsection{`Thermodynamic' equilibrium}
The combination of distributed sources presented above allows stable
steady-state solutions, which were analyzed in the framework of a
'closed-box' model \cite{kritsuk96}.  These solutions describe the ISM
in `thermodynamic' equilibrium, i.e. in a state, where the sources and
sinks of mass and energy are locally balanced.

An equilibrium temperature of one such state, calculated for our
assumed set of parameter values, is $T_{\rmn{eq}} (\alpha, b, T_0,Z)
\approx 1.4\times10^7$~K.  This temperature is close to the
temperature estimates for hot gaseous coronae based upon the X-ray
data of nearby giant ellipticals.
 
The characteristic time scale, on which the gas is able to reach
thermodynamic equilibrium from any point in the huge attraction region
of the phase plane $(\rho, T)$,
\begin{equation}
t_{\rmn{s}}=\sqrt{t_{\rmn{c}}t_{\alpha}}, \label{TimeScale}
\end{equation}
is the geometric mean of the local cooling time $t_{\rmn{c}} \equiv
c^2/[(\gamma-1) \rho \Lambda]$ and the time scale for stellar mass
loss $t_{\alpha} = (\alpha\rho_*/\rho)^{-1}$; $c$ is the isothermal
sound speed, and $\gamma={5\over3}$ is the ratio of specific heats.

\subsection{Hydrostatic equilibrium}
For simplicity we assume that the source parameters $\alpha$, $b$,
$T_0$, and $Z$ do not depend on radius.  This implies that the
equilibrium temperature $T_{\rmn{eq}}$ is also radius independent.
Hence, if the gas is in thermodynamic equilibrium, it can
hydrostatically fill only the potential well of an isothermal sphere.
It is known, however, that due to dissipation the shapes of 
density profiles of
stars and dark matter usually do not coincide in giant galaxies.  We
therefore choose a potential of a two-component isothermal sphere,
which incorporates stars and dark matter with different velocity
dispersions \cite{kritsuk96,kritsuk97}. We define the parameter
$\beta$ as a ratio of these dispersions,
\begin{equation}
\beta\equiv{\sigma_*^2\over\sigma_{\rmn{DM}}^2}<1.
\end{equation}
This two-component model is completely defined by four parameters.
Once, we fix $\sigma_{\rmn{DM}}^2 = {{\cal R}\over \mu} T_{\rmn{eq}}$
(since hot gas and dark matter are usually distributed similarly) and
$\beta$=0.5 [see \cite{kritsuk96} for details], we have to define only
two additional quantities, namely, the characteristic radius of the
gravitating matter distribution
\begin{equation}
r_0=\frac{\sigma_*}{\sqrt{4\pi G \rho_{*,0}}}, \label{king_rad}
\end{equation}
and the ratio of dark matter density to stellar density at the centre
\begin{equation}
\delta=\rho_{\rmn{DM},0}/\rho_{*,0}.
\end{equation}
We set $\delta=10^{-1.5}$, since the stellar component dominates by
mass in the central regions of giant ellipticals.  Such a choice of
$\beta$ and $\delta$, which control the shape of the density
distributions \cite{kritsuk97a}, corresponds to a galaxy, whose surface
brightness profile follows the de Vaucouleurs law, and which contains
comparable amounts of mass in luminous and dark matter within its
effective radius.  Finally, in order to fix the scaling of physical
values, we set the characteristic radius $r_0 = 0.16$~kpc.  This
particular choice of parameters allows us to reproduce with reasonable
accuracy the shape of the surface brightness profile for NGC~4472
(M49), the optically most luminous E2 galaxy in the Virgo cluster,
assuming a distance of 17~Mpc \cite{trinchieri..86}.

Numerical integration of the Poisson and Jeans equations for the given
set of parameters provides us with a giant spheroidal model galaxy
with a total mass in stars $M_* \approx 8.5\times10^{11}$~M$_{\odot}$
and a stellar velocity dispersion $\sigma_* \approx 304$~km~s$^{-1}$.

The hydrostatic isothermal gas distribution in its gravitational
potential follows the so-called $\beta$-law: $\rho_{\rmn{eq}} \propto
\rho_*^{\beta}$ with $\beta=0.5$.  This implies that, as it is
observed, the shapes of the optical and the X-ray brightness profiles
are identical [cf. \cite{trinchieri..86}] and $T_{\rmn{eq}}=2T_*$
[cf. \cite{davis.96}].  The self-gravity of the ISM is negligible and
is not taken into account.
 
The central equilibrium gas density is $\rho_{\rmn{eq,0}}/(\mu
m_{\rmn{H}})\approx0.19$~cm$^{-3}$ and the time scale associated with
the sources at the centre is $t_{\rmn{s,0}} \approx 4.9\times10^7$~yr.
The gravitational time scale
\begin{equation}
t_{\rmn{G}}={1\over\sqrt{4\pi G\rho_{*,0}}}=
{r_0\over\sigma_*}\approx2.1\times10^6\;\rmn{yr} 
\end{equation}
is much shorter than $t_{\rmn{s,0}}$.  The cooling radius, fixed at
the distance where $t_{\rmn{c}}=10$~Gyr, is $r_{\rmn{cool}}=36$~kpc.

\subsection{Basic equations}
We probe the global stability of the equilibrium hydrostatic gas
configuration with respect to a variety of perturbations by means of
numerical solutions of the time-dependent hydrodynamic equations for
the mass, momentum and energy balance in the hot ISM:
\begin{equation}
\frac{\partial \rho}{\partial t}+\nabla\cdot(\rho \bmath{v})=\alpha
\rho_* - \dot\rho_{\rmn{ti}}, \label{Mass}
\end{equation}
\begin{equation}
\frac{\partial (\rho \bmath{ v})}{\partial t}+\nabla(\rho
\bmath{v}^2+p)=\rho\nabla\phi - \dot\rho_{\rmn{ti}}
\bmath{v}, \label{CompleteSystem}
\end{equation}
\vspace{-5pt}
\begin{eqnarray}
\lefteqn{\frac{\partial E}{\partial t}+\nabla\cdot[\bmath{v}(E+p)] = } 
\nonumber \\ 
& & \alpha\rho_* e_* -\dot\rho_{\rmn{ti}}(E+p)/\rho-\rho^2\Lambda +\rho
\bmath{v}\cdot\nabla\phi-\nabla\cdot\bmath{q}. \label{Energy}
\end{eqnarray}
Here $\bmath{ v}$ is the velocity, $p=(\gamma-1)e\rho$ is the
pressure, $e$ is the specific internal energy, $e_* = c_{\rmn{v}}
T_0$, and the energy density $E=\rho(e+\bmath{v}^2/2)$.

\subsection{Initial conditions\label{initial}}
The undisturbed state is considered to be a spherically symmetric gas
configuration in hydrostatic and thermodynamic equilibrium:
\begin{eqnarray}
\left.\rho\right|_{t=0}&=& \rho_{\rmn{eq}}(r),\label{eq_1} \\
\left.p\right|_{t=0}&=& p_{\rmn{eq}}(r)\equiv {{\cal
R}\over\mu}\rho_{\rmn{eq}}T_{\rmn{eq}}, \label{eq_2} \\
\left.\bmath{v}\right|_{t=0}&=&0.\label{eq_3}
\end{eqnarray}
For one-dimensional spherically symmetric simulations we introduce a
global density perturbation with an amplitude $\epsilon=const$:
\begin{equation}
\left.\rho\right|_{t=0}=(1+\epsilon)\rho_{\rmn{eq}}(r). \label{init_1}
\end{equation}
For the 2D axisymmetric simulations, which were performed in spherical
coordinates $(r,\vartheta,\varphi)$, we used either a single-wave
isothermal density perturbation of the form
\begin{equation}
\left.\rho\right|_{t=0}=\left[1+\epsilon\left(\cos^2{\vartheta}
-0.5\right)\right]\rho_{\rmn{eq}}(r), 
\label{init_2} 
\end{equation}
or random large-scale velocity field perturbations, $\bmath{
v}=(u,v)$, (where $u$ and $v$ are the radial and the tangential
velocity component, respectively) which are defined on the discrete
grid (see section \ref{grid}) as follows:
\begin{equation}
\left.u,v\right|_{t=0}= \delta_{u,v}(r,\vartheta),
\end{equation}
\begin{eqnarray}\label{init_3}
\delta_a(r_k,\vartheta_l)&=&{1\over Q}\sum_{i=1}^3\sum_{j=1}^3
\left(A_{ij}^a\cos {2\pi\over n}ik \sin {2\pi\over n}jl + \right.\\
&&B_{ij}^a\cos {2\pi\over n}ik \cos {2\pi\over n}jl + \nonumber\\
&&C_{ij}^a\sin {2\pi\over n}ik \sin {2\pi\over n}jl + \nonumber\\
&&\left.D_{ij}^a\sin {2\pi\over n}ik \cos {2\pi\over n}jl\right), \nonumber 
\end{eqnarray}
with $a\in\{u,v\}$. The computational volume is discretized into
$n\times n$ cells, which are labeled by the pair of indices $k, l$.
$\{A,B,C,D\}^a_{ij}$ are quasi-random numbers from the interval
$[-1,\:1]$. The amplitude of the velocity perturbations is normalized
by a factor $Q$, so that ${1 \over 2}(\max_{k,l}\delta_a - \min_{k,l}
\delta_a) = \epsilon$.\footnote{Velocities are given in units of the
adiabatic sound speed $c(T_0)\approx812$~km~s$^{-1}$.}

We varied the amplitude $\epsilon$ in different experiments in the
range from 0.2\% to 30\%.

\subsection{Boundary conditions\label{boundary}}
We imposed the following conditions on the hydrodynamic variables at
the inner spherical boundary at $r_{\rmn{i}} \approx0.12\,$kpc:
\begin{equation}
u(r_{\rmn{i}})=0, \;\;\;\;\;\;\;\; \left.{\partial u\over \partial r}
\right|_{r=r_{\rmn{i}}}=0,\label{bound_i_B}
\end{equation}
\begin{equation}
v(r_{\rmn{i}})=0, \;\;\;\;\;\;\;\; \left.{\partial v\over \partial r}
\right|_{r=r_{\rmn{i}}}=0,
\end{equation}
\begin{equation}
\left.{\partial T\over \partial r}\right|_{r=r_{\rmn{i}}}=0,
\end{equation}
\begin{equation}
\left.{\partial p\over \partial r}\right|_{r=r_{\rmn{i}}}=\left.
\rho{\nabla\phi}\right|_{r=r_{\rmn{i}}}.\label{bound_i_E}
\end{equation}
At the outer spherical boundary at $r_{\rmn{o}}\approx151\,$kpc, we used
slightly different conditions:
\begin{equation}
u(r_{\rmn{o}})=0, \;\;\;\;\;\;\;\; \left.{\partial u\over \partial r}
\right|_{r=r_{\rmn{o}}}=0,
\end{equation}
\begin{equation}
v(r_{\rmn{o}})=0, \;\;\;\;\;\;\;\; \left.{\partial v\over \partial r}
\right|_{r=r_{\rmn{o}}}=0,
\end{equation}
\begin{equation}
T(r_{\rmn{o}})=T_{\rmn{eq}},
\;\;\;\;\;\;\;\; \left.{\partial T\over \partial r}\right|_{r=r_{\rmn{o}}}=0,
\end{equation}
\begin{equation}
p(r_{\rmn{o}})=p_{\rmn{eq}}(r_{\rmn{o}}),
\;\;\;\;\;\;\;\;\left.{\partial p\over \partial r}\right|_{r=
r_{\rmn{o}}}=\left.\rho{\nabla\phi}\right|_{r=r_{\rmn{o}}}.
\end{equation}
In the 2D axisymmetric calculations we solved the flow equations in a
90 degree sector centered at the equatorial plane and imposed
`periodic' boundary conditions in angular direction at
$\vartheta_{\rmn{l,r}} = {\pi\over2} \pm {\pi\over4}$:
\begin{equation}
a(\vartheta_{\rmn{l}})=a(\vartheta_{\rmn{r}}),
\;\;\;\;\;\;\;\;\left.{\partial a\over \partial 
\vartheta}\right|_{\vartheta=\vartheta_{\rmn{l}}}=
\left.{\partial a\over \partial \vartheta}\right|_{\vartheta=
\vartheta_{\rmn{r}}},
\end{equation}
where $a\in\{\rho,u,v,p\}$.

\section{Numerical method}
The numerical solutions discussed in this paper were obtained using
the {\em Piecewise Parabolic Method} (PPM) \cite{colella.84} -- a
second order extension of Godunov's finite difference scheme, which
uses a nonlinear Riemann solver in the hydrodynamic calculations.  It
was originally described in \cite{godunov59,godunov..61} and first
adapted for high order difference schemes by van Leer
\shortcite{vanleer79}.  We use a direct single-step explicit Eulerian
formulation of the method for the equations written in conservation
form. We solve the equations in spherical coordinates for both
spherically symmetric and axisymmetric problems.  The coordinate
singularities at the centre and along the symmetry axis are isolated
by a proper choice of the computational volume and the corresponding
boundary conditions (see above).

There are slight differences in our PPM implementation, as compared to
the original prescription of Colella \& Woodward
\shortcite{colella.84}. These are stipulated by the presence of source
terms, which play an essential role in our modeling.

\subsection{Treatment of the source terms}
In order to keep the numerical scheme consistent, the expressions for
$\beta^0$ and $\beta^{\pm}$ [see equation (3.7) of \cite{colella.84}]
have been re-derived to incorporate the terms corresponding to the
volume sources of mass, momentum and energy, due to radiative cooling,
stellar mass loss, SNe heating and local thermal instabilities.  This
gives rise to a modification of the procedure for the calculation of
the effective left and right states for the Riemann problem.  The
effect of the source terms on $\overline{U}_{j+1/2}$\footnote{We keep
here the notation of Colella \& Woodward \shortcite{colella.84}.} is
accounted for to second order accuracy.  The detailed formulae can be
found in \cite{kritsuk96a}.

The presence of sources, represented by non-linear functions of the
hydrodynamic variables, considerably modifies the behaviour of the gas
flow.  In particular, when the effects of local thermal instabilities
are taken into account, the elastic properties of the gas change
dramatically.  In simple terms, slow gas compression by an external
force easily results in condensation of a significant fraction of the
gas. The self-accelerating contraction does not meet any considerable
resistance by pressure forces, as long as it is slow enough.  When
simulating such gas flows we observed the formation of `condensation
fronts' -- thin sheet-like quasi-isobaric structures, characterized by
an enhanced density and negative velocity divergence.  These fronts,
being the major sites of condensation, are able to move in the
direction opposite to large-scale pressure gradients and coalesce with
each other upon collisions.  In order to treat this kind of solutions
properly with our numerical scheme, we had to introduce additional
dissipation in the vicinity of the fronts.  The dissipation algorithm
allows us to keep the peak density values finite, limiting the
run-away cooling and condensation of the gas in cells located at the
front.  In fact, we added an explicit diffusive flux to the numerical
fluxes in the mass and momentum conservation equations following the
simple prescription of Colella \& Woodward (1984).  Further details of
the dissipation algorithm can be found in \cite{kritsuk96a}.  While we
apply the simple flattening around shock fronts [see the Appendix in
\cite{colella.84}] in our 1D calculations, we have switched it off in
our 2D simulations.

The sources make the method implicit, since the values of $\rho^{n+1}$
and $\rho^{n+1}u^{n+1}$ depend on $E^{n+1}$ through terms, which
include temperature dependent radiative cooling $\Lambda$ and mass
deposition $\dot\rho_{\rmn{ti}}$.  To avoid an iterative solution
method we approximated the unknown values at $t = t^{n+1}$ by those
from the previous time level at $t = t^n$ in the source terms
$\dot\rho_{\rmn{ti}}^{n+1}$ and $\Lambda^{n+1}$, still using the
correct treatment elsewhere.

\subsection{Introducing the heat flux\label{heat}}
In order to estimate the stabilizing effects of thermal conductivity
on the isothermal equilibrium solution we include a $\nabla \cdot
\boldmath{q}$ term in the final conservative difference step.  To keep
the scheme explicit, we estimate the temperature values at zone
interfaces using the computed left and right state values for pressure
and density.  Linear approximations of the temperature gradients at
the interfaces are computed using the temperature differences in
adjacent zones.  An adequate reduction of the time step was
implemented, too.  Such a simplified treatment for the heat conduction
flux still allows us to keep the accuracy of the computations
sufficiently high in a broad vicinity of the equilibrium solution.

\subsection{The computational grid \label{grid}}
The computational grid is organized in such a way that it allows us to
describe the physically discrete sources as continuously distributed in
space.  In our particular model the largest scale associated with an
individual source is the scale for local thermal instability 
($l^{\rmn{ti}}>l^{\rmn{sn}}>l^{\rmn{sw}}$).  This
scale depends on the degree of suppression of heat conduction by
magnetic fields and on the spectrum of the primeval perturbations,
which are supplied by ongoing evolutionary processes in the galaxy.
\begin{figure*}
\vspace{-5cm}
\psfig{figure=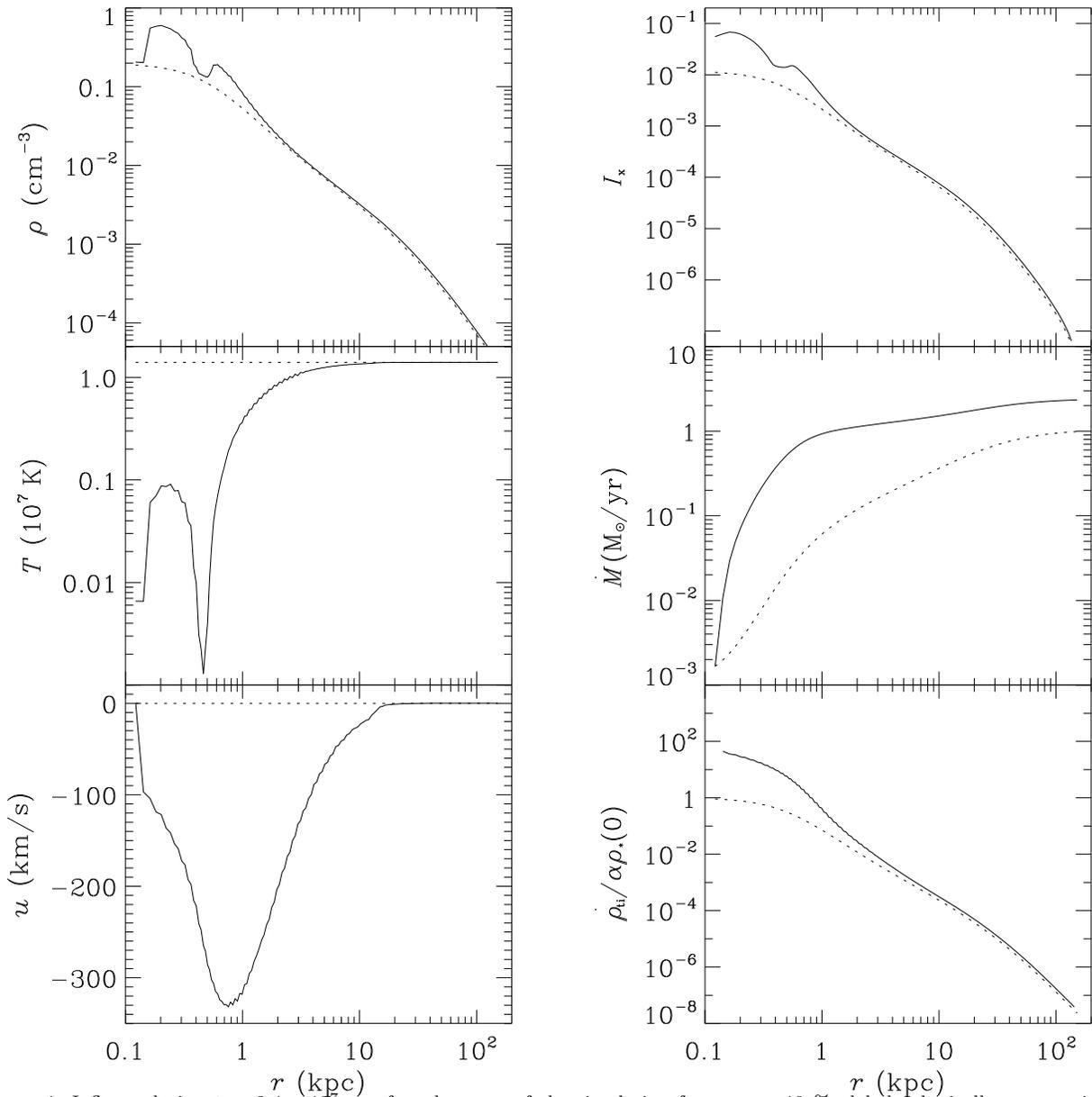,clip=}
\vspace{-4.3cm}
\caption{Inflow solution $t=2.4\times10^7$~yr after the start of the
simulation for an $\epsilon=10$~\% global spherically symmetric
density perturbation (solid lines).  Dotted lines indicate the
unperturbed hydrostatic equilibrium solution.  The plots show density,
temperature and radial velocity of the gas (left column) together with
the surface brightness in X-rays, $I_{\rmn{x}}$, the mass deposition
rate and the efficiency of mass dropout due to thermal instabilities
(right column) as functions of the radius.  The surface brightness is
given in arbitrary units, see text for details.}
\label{inflow}
\end{figure*}
We introduce a correction factor $\eta$, which defines the scale for
growing local thermal instabilities 
$l^{\rmn{ti}} \sim 0.3\:\eta\:T_7^2/n_{-1}$ kpc.  With
$\eta \approx 0.1$, in the core region of the gas distribution
$l^{\rmn{ti}} \sim 0.03\,$kpc becomes smaller than the
minimum scale on which supernovae contribute as a continuously distributed
 source,
$l^{\rmn{sn}} \sim 0.06\,$kpc for the assumed SN rate
[see \cite{kritsuk92} and references therein for more details].  Thus,
if the grid cells are larger than $\approx 10^{-2}\,$kpc in the core
region, the sources can be treated as being continuously distributed.

The grid is organized as follows.  The inner $k$ radial cells are
equidistant: $r_j = r_{j-1} + \Delta_{\rmn{r}}$ for $j=1, \ldots,k-1$,
starting from $r_{\rmn{i}} \approx 0.12\,$kpc, where $\Delta_{\rmn{r}}
\approx 0.020\,$kpc, and $k=20$. Hence, the core region of the initial
density distribution (King's core radius is $r_{\rmn{c}}\approx 3
r_0$) is covered by a uniform radial grid.  At larger radii an
exponential grid spacing is used: $r_j = r_{j-1} \Delta_{\rmn{r}}
\exp\left(6.2{j-k\over m-k}\right)$ for $j=k,\ldots,m-1$.  This
spacing guarantees a smooth transition from a uniformly to a
non-uniformly spaced grid at $r \approx 0.40\,$kpc.  The total number
of radial grid points is $m=128$.

In the 2D simulations we used the same $r$-spacing and an equidistant
angular grid of 128 zones with $\Delta_{\vartheta} = \pi/2/128$.

\section{Evolution of spherically symmetric perturbations}

\begin{figure*}
\vspace{-5cm}
\psfig{figure=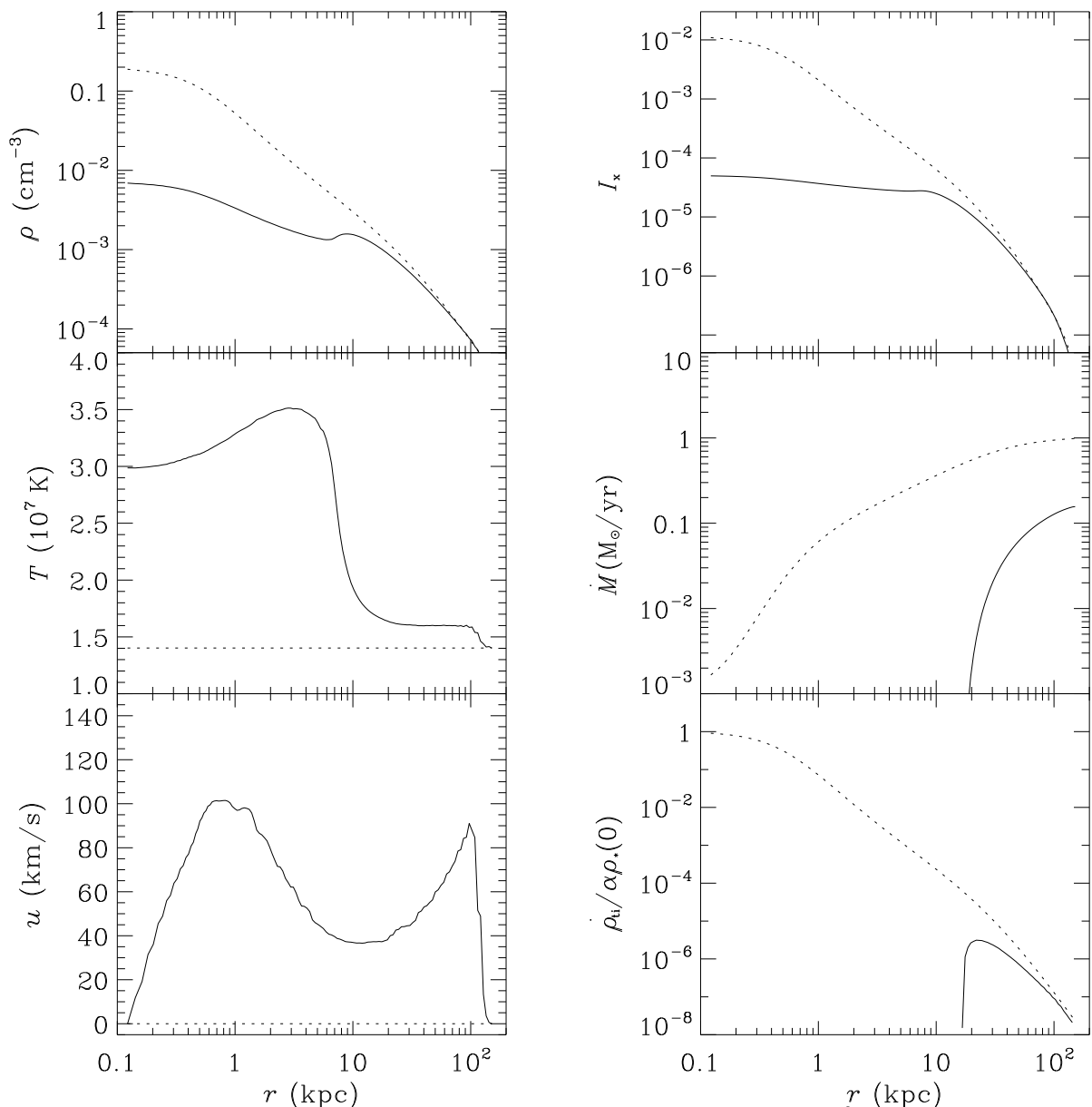,clip=}
\vspace{-4.3cm}
\caption{ An example of a spherically symmetric outflow solution at
$t=2.0\times10^8$~yr; see also Fig. 1.}
\label{outflow}
\end{figure*}

\subsection{Case of suppressed heat conduction}
First, we discuss the global stability of the hydrostatic equilibrium
gas configuration assuming that heat conduction is completely
suppressed on large scales.

Our simulations clearly demonstrate that in the absence of efficient
heat conduction the equilibrium is unstable.  For example, in response
to a 10\% positive density perturbation of the form (\ref{init_1}) an
inflow develops on a central cooling time-scale $t_{\rmn{c,0}} \approx
0.5\,t_{\rmn{s},0}$ (see Fig.~\ref{inflow}), whereas the same negative
density perturbation initiates an outflow (see Fig.~\ref{outflow}).

With our numerical scheme we are not able to describe the hydrostatic
equilibrium of the initial gas configuration perfectly.  Consequently,
the `unperturbed' initial model generates some wave motions, which
help the gas to settle in the potential well.  This provides some
initial heating, i.e. the `unperturbed' model generates an outflow.
The minimum amplitude which is necessary to overcome the initial noise
inherent in the numerical approximation of the equilibrium model is 
$\epsilon \approx 0.0045$ for
spherically symmetric density perturbations

\subsubsection{Inflow solutions}
A positive gas density perturbation first breaks the balance between
cooling and heating in the core region of the gas distribution.
Gravity immediately starts to accelerate the gas inflow.  As a result,
a dense nearly isothermal gas core with $T \approx 10^6$~K forms,
which is condensationally unstable.  The further development of
instability depends on the detailed structure of the inflow velocity
profile.  The condensation occurs first at a radius, where
$\nabla\cdot\bmath{v}$ has a local minimum.  In the example shown in
Fig.~\ref{inflow}, this happens nearly simultaneously at the inner
boundary and at $r \approx0.5\,$kpc.  Two sharp minima in the
temperature profile (with $T \approx10^4\,$K) and depressions in
density distribution due to the high condensation efficiency can be
clearly seen. For $\eta\in[0,\:0.1]$ the mass loss rate
\begin{equation}
\dot M(r)\equiv4\pi\int_0^{r}r^2\dot\rho_{\rmn{ti}}{\rmn{d}}r
\label{mdot}
\end{equation}
at this transient stage is $2-4$~M$_{\odot}$~yr$^{-1}$ within
$r_{\rmn{cool}}$. Most of the condensate is deposited inside a radius
of $1-2\,$kpc.  The surface brightness
\footnote{Here we conventionally call `surface brightness' the
emission measure of the unit column along the line of sight. The X-ray
surface brightness includes comparable contributions from both
$\rho^2\Lambda$ and $\dot\rho_{\rmn{ti}}e$, so that the combined
spectral volume emissivity of the cooling hot gas and local unstable
condensates is $\epsilon_{\nu}(r) = \rho^2(r) \Lambda_{\nu}(T) + \dot
\rho_{\rmn{ti}}(r) c_{\rmn{v}} \Gamma_\nu(T)$ \cite{kritsuk96}.  This
can be rewritten in the form $\epsilon_{\nu}(r) = \rho^2(r) f_1(\nu,
T) - \rho_{\rmn{eq}}^2 f_2(\nu, T)$, where $f_{1,2}$ are functions of
temperature and frequency.  In order to get an idea of the shape of
the surface brightness profile, which can be observed, we define
$I_{\rmn{X}} \equiv {2\over r_0} \int_R^{\infty}{\rho^2r{\rmn{d}}r
\over \rho_{\rmn{eq},0}^2 \sqrt{r^2-R^2}}$.  One has to be cautious,
however, recovering the surface brightness profiles from the emission
measure in essentially non-isothermal flows.}
in the core region grows by a factor of ~10 over its equilibrium
value.  We had to terminate the calculation at $t = 0.49\,
t_{\rmn{s,0}} \approx 2.4\times10^7$~yr because the iteration within
Riemann solver no longer converged when the temperature drops below
$10^4$~K.

Such accelerated, non-linear unstable inflow regimes can obviously
exist only in configurations with perfectly supported spherical
symmetry.  This is an unlikely situation for real galaxies, and hence
points towards the need for multidimensional simulations.

The existence of an exact hydrostatic equilibrium solution allows us
to define the boundary conditions properly.  Conditions at the inner
boundary are of particular importance for the case of inflow
solutions.  Free boundary conditions (i.e. zero spatial derivatives of
all dependent variables) are inconsistent with hydrostatic initial
conditions, because the gas is forced to flow towards the centre,
having no hydrostatic support at the inner boundary.  Our choice [see
equations (\ref{bound_i_B})--(\ref{bound_i_E})] seems to be optimal
for this particular problem.  In case of a fixed temperature at its
equilibrium value, $T\left.\right|_{r=r_{\rmn{i}}} = T_{\rmn{eq}}$,
spurious oscillations are generated near the inner boundary which
accelerate the development of the central cooling catastrophe
considerably.

A kind of settling `cooling flow' solution was found by David et
al. (1990) \nocite{david..90} in time-dependent simulations of hot gas
flows in elliptical galaxies.  They ignored thermal conductivity and
included both, sinks and sources for the gas mass, although with a
slightly different parameterization.  Apparently, the existence of
these settling solutions is mainly due to their specific choice of the
inner boundary conditions.  They use a King profile for the stellar
mass distribution and keep the derivatives of temperature and density
equal to zero at the inner boundary, located in the core region,
allowing for a free gas inflow through the boundary towards the
galactic centre.  This simply cuts out the central region, where the
cooling catastrophe would have developed.

\subsubsection{Outflow solutions}
The slow outflow solution describes the formation of a hot,
subsonically expanding bubble filling the central $r\ge10$~kpc of the
galaxy (see Fig.~\ref{outflow}).  The gas density gradually drops
below the equilibrium values in the region covered by the outflow,
while its temperature rises up to $T\approx4\times10^7$~K.  Gas
deposition due to thermal instabilities in the overheated and rarefied
plasma becomes inefficient, and the gas shed by stars starts to blow
out from the centre.  As a result, a characteristic feature of cooling
flows -- a strong central peak of the X-ray surface brightness -- is
being smeared out and the X-ray luminosity of the galaxy decreases
dramatically.  We terminated the calculation at $t \approx
4\:t_{\rmn{s,0}}$, when the leading wave front of the outflow region
reaches the outer boundary at $r_{\rmn{o}}=151$~kpc.

Spherically symmetric outflows develop in a very stable, gradual way
and, obviously, can serve as an explanation for the low
X-ray-to-optical luminosity ratios $L_{\rmn{X}}/L_{\rmn{B}}$ of some
ellipticals.  In our case they originate in a natural way from an
equilibrium gas configuration and do not require any coordinated
temporal changes of the SN rate and/or stellar mass loss rate,
cf. \cite{ciotti...91}.

\subsection{Stabilizing the equilibrium by thermal conductivity}
In order to check the effects of the conductive heat flux, we made a
number of runs, varying the reduction factor $\eta$ and the
perturbation amplitude $\epsilon$ in a wide range of values.

\begin{table}
\caption{Stabilizing effect of the conductive heat flux$^a$}
\begin{center}
\begin{tabular}{lcccccc}
\hline\hline
 & \multicolumn{3}{c}{inflow: $t_{\rmn{cc}}/t_{\rmn{s,0}}$}
 &\multicolumn{3}{c}{outflow: $u_{\rmn{max}}$~[km~s$^{-1}$]}
\\ \cline{2-4}\cline{5-7}
$\pm\epsilon$&
$\eta=0$&$\eta=0.3$&$\eta=1$&$\eta=0$&$\eta=0.3$&$\eta=1$\\
$0.005$&1.59&$\infty$&$\infty$&196   & 20.5 &2.3 \\
$0.01 $&1.08&1.76    &$\infty$&197   & 21.3 &2.8 \\
$0.1  $&0.49&0.59    &1.33    &241   & 32.5 &10.6\\ \hline
\end{tabular}
\medskip
\end{center}
\medskip
{\small $^a$ The time-scale $t_{\rmn{cc}}$ for the development of the
cooling catastrophe is given in units of the characteristic time for
sources at the centre, $t_{\rmn{s,0}}$.  Infinity signs indicate that
the cooling catastrophe does not develop. In this case the solutions
either show an oscillatory behaviour around the hydrostatic one or
evolve into a very subsonic outflow.  The values of $u_{\rmn{max}}$
give the maximum velocities reached by the flow during the evolution
until $t=10\:t_{\rmn{s,0}}$.  The initial perturbation amplitude
$\epsilon$ is positive in case of inflow solutions and negative for
outflows.}
\end{table}

\begin{figure*}
\vspace{-5cm}
\psfig{figure=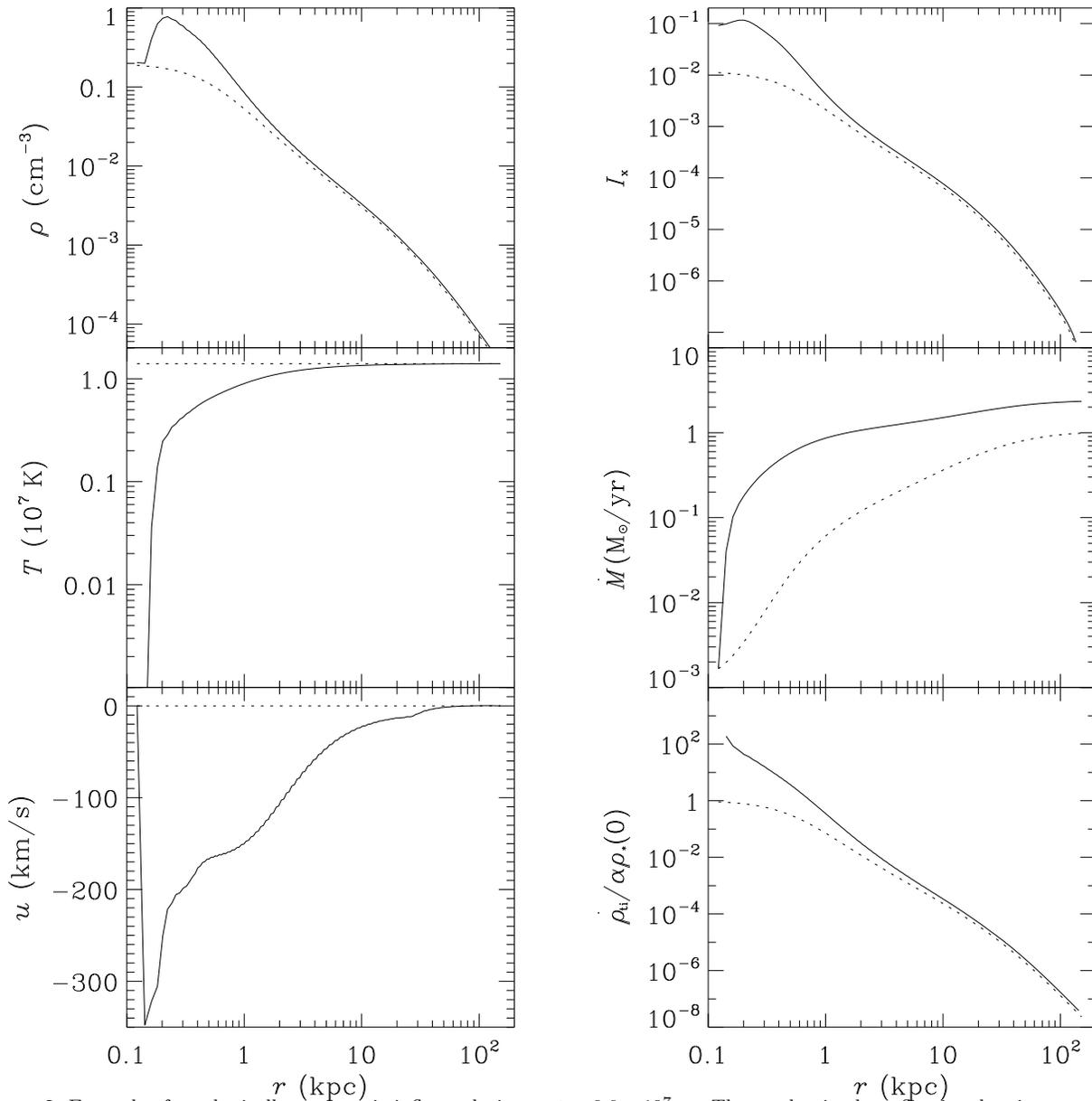,clip=}
\vspace{-4.3cm}
\caption{Example of a spherically symmetric inflow solution at
$t=6.6\times10^7$~yr. The conductive heat flux is taken into account
with $\eta=1$. See also Fig.~1.}
\label{inflow_cond}
\end{figure*}

\begin{figure*}
\vspace{-5cm}
\psfig{figure=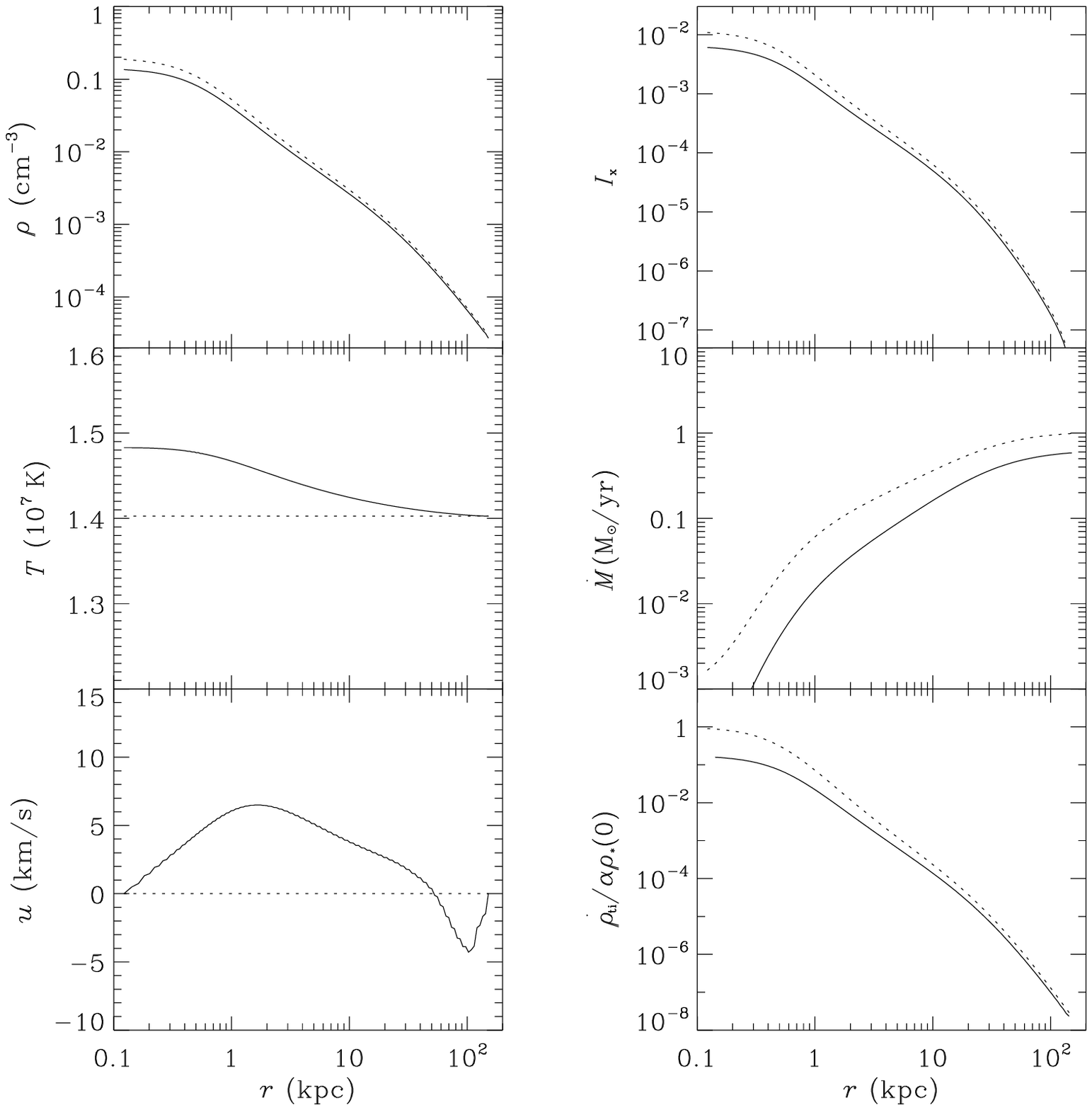,clip=}
\vspace{-4.3cm}
\caption{Example of a steady-state slow outflow solution at
$t=4.9\times10^8$~yr. The conductive heat flux is taken into account
with $\eta=1$. See also Fig.~1.}
\label{outflow_cond}
\end{figure*}

The data given in Table~1 show that thermal conductivity is more
effective in damping the non-linear development of negative density
perturbations than those of positive ones.  It efficiently suppresses
the development of outflows and keeps the gas velocities quite low
even for quite large initial perturbations (see
Fig. \ref{outflow_cond}).  It also affects the development of positive
density perturbations by delaying the onset of the cooling catastrophe
and by creating a negative temperature gradient in the core region
which shifts the onset of condensation towards the inner boundary (see
Fig. \ref{inflow_cond}).
We find, however, that a heat flux is able to prevent catastrophic
cooling only, if it is not suppressed and if the initial perturbations
are sufficiently small.  Solutions obtained in runs with $\eta=1$ and
$\epsilon=0.5\%$ or 1\% exhibit a stable quasi-hydrostatic behaviour
at $t \approx 10\:t_{\rmn{s,0}}$ with deviations from the equilibrium
temperature of the order of 1\% and inflow/outflow velocities
$u\le1$~km~s$^{-1}$.

We further find that even a considerably suppressed heat flux damps
the small-scale numerical noise, which one can infer from the
temperature panel in Fig. \ref{inflow}.  This noise is amplified by a
physical instability similar to that occurring in Field's instability.
If $\eta=0$, the noise saturates at a level of the order of 1\%.

The results of these simulations confirm the
earlier qualitative finding of Meiksin (1988) \nocite{meiksin88}, who
showed that heat conduction delays the development of the cooling core,
but does not prevent it (see Table 1).  However, in contrast to
\cite{meiksin88}, we did not find any solution of the `cooling flow'
type with a considerable inflow velocity, which is reasonably stable.
We attribute this difference mainly to stellar mass loss, which was
ignored in \cite{meiksin88}.  It provides a steady mass supply and
simultaneously stabilizes local thermal instabilities by reducing the
sink terms [see equation (\ref{GrowthRate})].  Both effects amplify
the cooling and destabilize the flow near the centre.  Since central
galaxies do always reside in cooling flows, stable inflow solutions of
this type seem to be unrealistic.

Instead, our results demonstrate that steady-state slow outflow
solutions of the type shown in Fig. \ref{outflow_cond} should be
common in ellipticals, since they are more stable than any of the
inflow solutions.  Their observable properties are close to those of
the equilibrium model, while the mass deposition rate is lower, $\dot
M(r_{\rmn{cool}}) \sim 0.4$~M$_{\odot}$~yr$^{-1}$ within the cooling
radius.  At the same time, the slow outflow solutions do not show a
central temperature depression.

\section{Axisymmetric perturbations}
If a spherically symmetric hydrostatic equilibrium gas configuration
is unstable it evolves either into inflow or into outflow. If one
relaxes the assumption of spherical symmetry and allows for 2D
perturbations and flow pattern, one can get hybrid solutions which
show both types of unstable behaviour in one model.  Our axisymmetric
simulations demonstrate that this is indeed the case.

We simulated the evolution of axisymmetric perturbations of various
kinds in a `toroidal' volume centered around the equatorial plane (see
sections \ref{initial}, \ref{boundary} and \ref{grid} for more
details).

An isothermal density perturbation of the form (\ref{init_2}) evolves
into a regular stream of cooling gas towards the centre in the
equatorial plane and its close vicinity.  The stream is embedded into
an extended slow outflow region covering the rest of the computational
volume.  The development of the instability occurs in several stages.
First, the gas density starts to grow in the perturbed region near the
centre.  Then, at some point ($t\approx0.5\:t_{\rmn{s,0}}$), gas
starts to cool and condense very efficiently in the inner part of the
equatorial plane ($r \le 3$ kpc) where the perturbations grow
non-linear first.  This produces a deep and narrow pressure (density)
minimum sandwiched by sharp pressure (density) maxima above and below
the equatorial plane, which smoothly approach the equilibrium pressure
(density) values near the angular grid boundaries.  The whole
configuration works as a `funnel' (hereafter in a 2D sense) which
sucks in rarefied gas from the periphery of the stratified hydrostatic
distribution towards the galactic centre.  Close to the inner
spherical boundary the highly accelerated gas stream passes through a
shock front and quickly condenses out in the dense post-shock region.

\begin{figure}
\psfig{figure=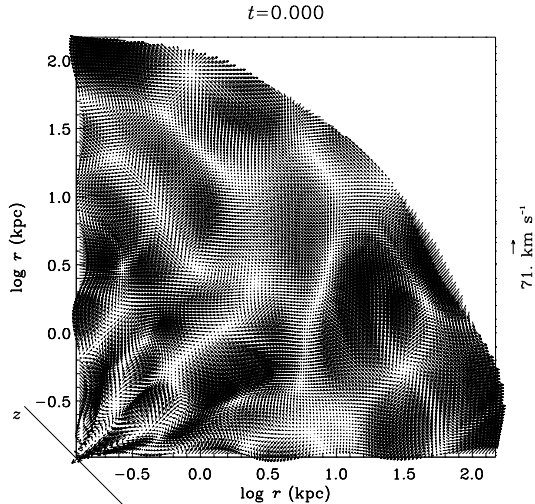,clip=}
\vspace{-0.3cm}
\caption{Example of an axisymmetric random velocity perturbation of
the form (\ref{init_3}) with an amplitude $\epsilon=10$~\%. The
straight line marked with $z$ indicates the symmetry axis.}
\label{velo0}
\end{figure}

\begin{figure}
\psfig{figure=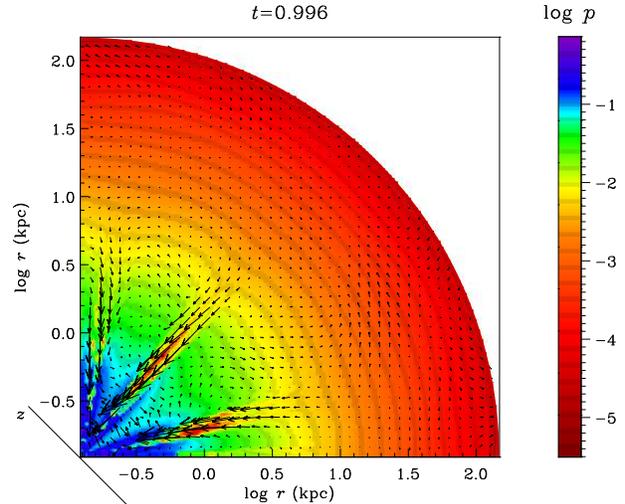,clip=}
\vspace{-0.3cm}
\caption{Example of an axisymmetric hybrid inflow-outflow solution.
Displayed are the distribution of gas pressure (colour map) and the
velocity field (arrows) at $4.9\times10^7$~yr, which forms in response
to the random velocity perturbation shown in Fig.~5.  The colour bar
labels give $p$ on a logarithmic scale in arbitrary units.}
\label{pressure}
\end{figure}

\begin{figure}
\psfig{figure=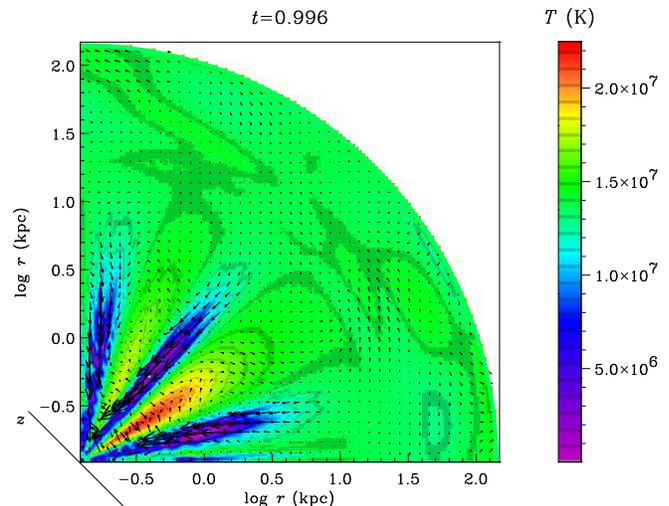,clip=}
\vspace{-0.3cm}
\caption{Same as Fig. 6, but showing the gas temperature (on a linear
scale).}
\label{temperature}
\end{figure}

\begin{figure}
\psfig{figure=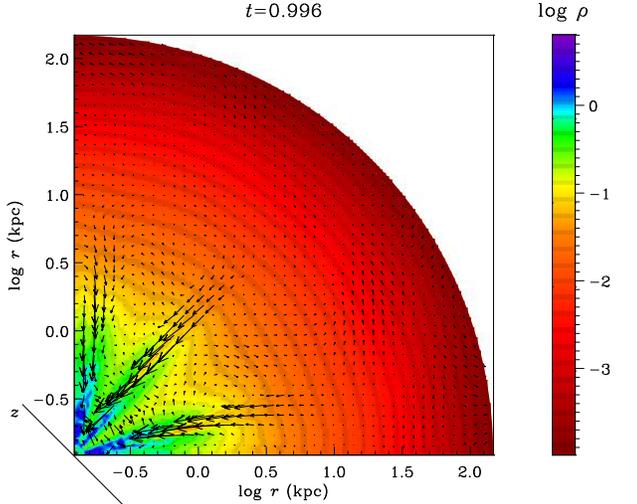,clip=}
\vspace{-0.3cm}
\caption{Same as Fig. 6, but showing the logarithm of the gas number
density in units of 0.28~cm$^{-3}$.}
\label{density} 
\end{figure}

\begin{figure}
\psfig{figure=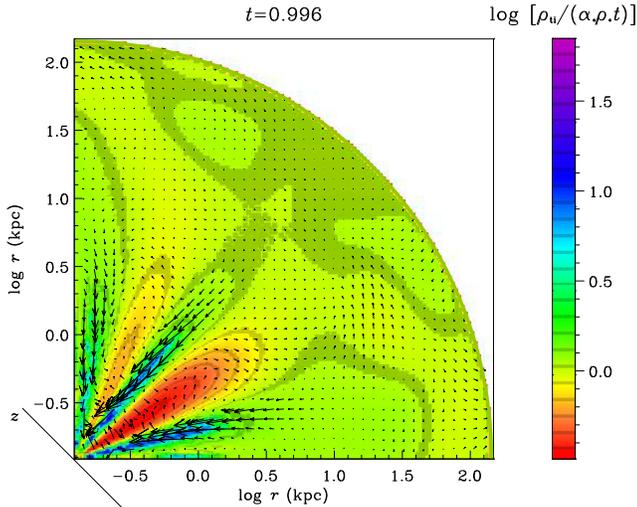,clip=}
\vspace{-0.3cm}
\caption{Same as Fig. 6, but showing the logarithm of the normalized
deposited gas density.}
\label{sink}
\end{figure}

\begin{figure}
\psfig{figure=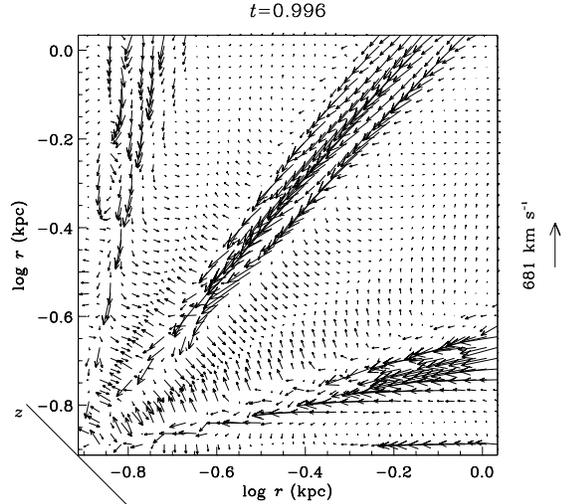,clip=}
\vspace{-0.3cm}
\caption{Velocity field in the central region at $t = 4.9 \times
10^7$~yr.}
\label{velocity}
\end{figure}

\begin{figure*}
\vspace{-5cm}
\psfig{figure=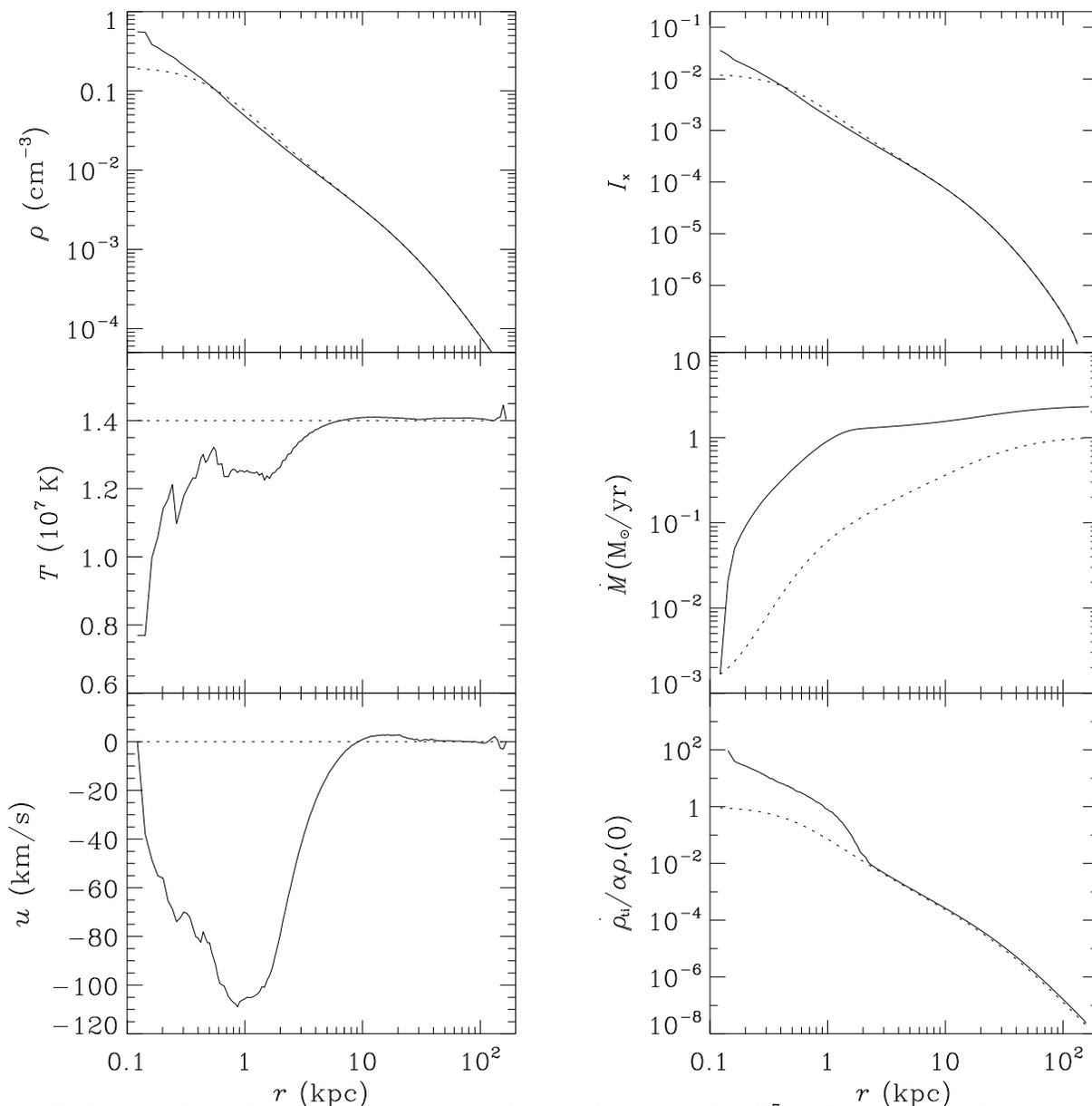,clip=}
\vspace{-4.3cm}
\caption{Example of an axisymmetric hybrid inflow-outflow solution at
$t = 4.9\times10^7$~yr. Displayed are various angular-averaged
variables. See also Fig. 1.}
\label{hybrid_av}
\end{figure*}

In order to model the processes in the hot galactic coronae more
realistically, we superimposed random perturbations of the velocity
field (\ref{init_3}) onto the equilibrium hydrostatic gas distribution
(\ref{eq_1})-(\ref{eq_3}); see Fig. \ref{velo0}.  Since these velocity
disturbances generate isobaric perturbations (which are then being
caught by the instability) of considerably smaller amplitude, it takes
longer to develop a non-linear flow.  We find that the flow enters the
non-linear regime at $t \approx t_{\rmn{s,0}}$. The results at $t =
1.0\:t_{\rmn{s,0}} \approx 5.0\times10^7$~yr are presented in Figs
\ref{pressure}-\ref{hybrid_av}.

The pressure map (Fig. \ref{pressure}) shows three funnels in the
central region ($r\le7\,$kpc).  The number of developed funnels and
their distribution in angular direction directly reflects the number
of harmonics present in the initial velocity perturbation and their
phases.  
Non-linear coupling of perturbation modes gives rise to a different
flow development in the individual funnels at $t = t_{\rmn{s,0}}$. 
In clockwise order, weak, strong, and medium
streams can be seen.

The arrows superimposed on the map give only a rough representation of
the flow pattern, because the ``arrow grid'' is sparser than the
computational grid to avoid illegible figures.  The velocity field for
the central 1~kpc region is shown in more detail in
Fig. \ref{velocity}.  Three streams of hot gas flow into the low
pressure regions located inside the funnels. Simultaneously, some gas
leaves the high pressure regions associated with the `funnel walls'
and enters the slow outflow regime between the funnels.

The pressure distribution in the leftmost funnel is rather patchy. The
corresponding features can also be seen in the velocity map
(Fig.~\ref{velocity}).  The gas tends to concentrate in radially
compressed clumps, where it condenses out digging a low pressure
channel into the hot galactic atmosphere.  As a result a radial hollow
jet of high velocity gas forms.  Growing from linear values of about
100~km~s$^{-1}$ at $t = 0.85\:t_{\rmn{s,0}}$ the maximum flow
velocity saturates at a slightly supersonic [in relation to
$c(T_{\rmn{eq}})$] value of $\le700$~km~s$^{-1}$ when the temperature
in the funnels drops to $\sim10^6$~K.

The inflow-outflow boundary represents a tangential discontinuity.  In
an ideal fluid such discontinuities are absolutely unstable even with
respect to infinitely small perturbations \cite{landau.59}.  The
growth rate of this shear instability, ${\rmn{Im}}
(\omega)$, depends on the wavenumber of the instability, the relative
velocity $v_{\rmn{r}}$ of the fluids and on the densities of both
fluids
\begin{equation}
{\rmn{Im}}(\omega)=kv_{\rmn{r}}{\sqrt{\rho_1
\rho_2}\over\rho_1+\rho_2}. 
\end{equation} 
On time-scales of $\sim0.1\:t_{\rmn{s,0}}$ perturbations with a
wavelength
\begin{equation}
\lambda\le \lambda_{\rmn{crit}}=0.6 {v_{\rmn{r}}\over100\: \rmn{km\:s}^{-1}}
{\sqrt{\delta\rho}\over1+\delta\rho}\;\:{\rmn{kpc}},\label{tang}
\end{equation} 
where $\delta\rho\equiv{\rho_1\over\rho_2}$, become non-linear.
However, our case is considerably more complex, since in the
non-linear regime compressibility effects are important and since the
instability of the tangential discontinuity can be coupled with the
condensational instability driven by the sources. Nevertheless, one
can still obtain a rough estimate from equation (\ref{tang}).  For a
typical density contrast $\delta\rho = 0.1$ and a relative flow speed
$v_{\rmn{r}} = 400$~km~s$^{-1}$ one finds $\lambda_{\rmn{crit}} =
0.7$~kpc.  This implies that the instability may be important for the
further evolution of the streams.

The temperature map (Fig. \ref{temperature}) shows that the initial
isothermal gas distribution gets destroyed by the perturbations.  The
funnels are filled by cooler gas with temperatures as low as $10^6$~K,
while in the outflow regions the gas is hotter than in the equilibrium
($T \approx 2\times 10^7$~K).  Note, that although periodic boundary
conditions are applied in angular direction, there exists no high
temperature region between the leftmost and (the periodically
continued) rightmost funnel.  Instead, there are indications of a
temperature drop and inflow in the inter-funnel region.  This may be
considered as an indication of an imminent coalescence of the two
evolved funnels, because of the same reason as in the case of the
coalescence of condensation fronts described in \cite{kritsuk96a}.
However, in the realistic three-dimensional case the probability for
such a coalescence of two finger-like streams must be quite low.  It
seems that at least the hot, well structured outflow region between
the central and the rightmost funnel will prevent them from merging.

The (number) density map (Fig. \ref{density}) looks qualitatively
similar to the pressure map.  Different stages of gas compression and
deposition can be seen in the three funnels.  The maximum number
density is 1.7~cm$^{-3}$. Such large values are reached in the
condensing 
fragments falling down towards the centre.\footnote{Note, 
that this is the density of the hot, pressure
supported gas phase rather than the density in deposited condensates,
which is much higher.}

The map of the normalized mass density (Fig.~\ref{sink}) deposited
during the simulation is similar to the temperature map.  Gas
condensation was much more efficient in the funnels, while in the
outflow regions the rate falls below the equilibrium values.  The
condensation pattern in the forming streams looks rather patchy.  This
indicates once again that during the development of the streams the
gas cools and condenses out in clumps.  With an initial number density
of 1.0 cm$^{-3}$ and a size of 0.2 kpc the mass of such a clump, $\sim
10^6$~M$_{\odot}$, corresponds to the mass of a globular cluster.

Apparently, the presence of hollow jets, which fill only a small
fraction of the whole volume in the central parts of a galaxy, does
not change the X-ray observables dramatically as compared to those of
the equilibrium gas configuration, if individual streams cannot be
resolved well.  To demonstrate this we show in Fig.~\ref{hybrid_av}
angular averages of the same variables, which were plotted for
spherically symmetric flows in Figs \ref{inflow}-\ref{outflow_cond}.
The mean temperature is averaged with a weight $\rho^2$, while we use
a $\rho$-weighted mean for the radial velocity.  We find that, at
least at $t = 1.0\:t_{\rmn{s,0}}$, the density and surface brightness
profiles do not differ substantially from the equilibrium ones.  The
mean temperature drops by $\sim50$~\% towards the centre showing some
oscillations due to individual cooling clumps.  Weighted mean inflow
velocities up to $\sim 10^2$~km~s$^{-1}$ indicate substantial
deviations from hydrostatic equilibrium.  [The adiabatic sound
velocity at the equilibrium temperature is $c(T_{\rmn{eq}}) \approx
555$~km~s$^{-1}$.]  At the same time, the bulk ($\sim50\%$) of the
condensed gas mass is deposited in the very center of the galaxy,
$r\le2$~kpc, and the mass deposition rate within the cooling radius is
comparable to that of the spherically symmetric inflows shown in Figs
\ref{inflow} and \ref{inflow_cond}.

These results suggest that in the absence of direct gas velocity
measurements the interpretation of the surface brightness and the
temperature profile in terms of the quasi-hydrostatic steady-state
approximation built into the `standard' cooling flow model can easily
result in an erroneous spatial distribution of the mass deposition
rate $\dot M(r)$.  Violation of hydrostatic equilibrium in a small
fraction of the volume, filled with accelerated gas streams, can lead
to a considerable mass flux of gas towards the centre, while the bulk
of the hot galactic corona remains in a quasi-equilibrium state.

Because of the extremely short cooling time scale in the blobs near
the centre, we cannot simulate the development of the instability into
the non-linear regime for a long time.  The final fate of the evolving
perturbations, thus, remains unclear at this stage of the study.  It
is hard to predict whether the formation of funnels represents just
the onset of a forthcoming global cooling catastrophe at the centre or
whether the funnels are transient features, which will be destroyed by
hydrodynamic instabilities.  Feedback heating could play an important
role in saturating the instability in the non-linear regime, but the
definition of feedback is uncertain.

\section{Discussion}
We found solutions, which can be classified in four different types:
\begin{itemize}
\item[(i)] A subsonic outflow with temperature enhancement in the
central region ${\Delta T\over T}\sim 3$ with typical gas velocities
of about 200 km~s$^{-1}$ and a low X-ray luminosity;
\item[(ii)] A quasi-hydrostatic gas configuration with outflow
velocities $\sim10$~km~s$^{-1}$, a nearly isothermal gas distribution,
and X-ray observables close to those of the recycling model;
\item[(iii)] An unstable spherically symmetric inflow with a X-ray
luminosity considerably exceeding observed values;
\item[(iv)] A hybrid inflow-outflow solution with a temperature
depression in the center, mean inflow velocities of $\sim
100$~km~s$^{-1}$, and a X-ray brightness distribution similar to the one
of the recycling model.
\end{itemize}
The prototypes are given in Figs \ref{outflow}, \ref{outflow_cond},
\ref{inflow}$+$\ref{inflow_cond}, and \ref{hybrid_av}, respectively.

Since the spherically symmetric non-linear regime of catastrophic
cooling is unstable, we consider solutions of type (iii) as being not
realized in nature.  The other three types represent flow regimes,
which can be expected to occur in galaxies,
\footnote{Note, that among these types there is no stable spherically
symmetric `cooling flow' realization with a subsonic inflow at all
radii.}
depending on the degree of suppression of thermal conductivity in the
hot ISM and on the source of the large-scale perturbations.

According to a simple scenario, the hot ISM evolves through several
stages with a galaxy free of any gas in the beginning: (1) a
supersonic galactic wind driven by early SN activity, (2) a slow
outflow regime, and (3) an inflow, triggered by the central cooling
catastrophe; see \cite{ciotti...91}.  The evolution is driven by
coordinated secular changes of the control parameters which determine
the mass loss rate and the characteristic temperature of the
distributed heat source in the galaxy, $\alpha(t)$ and $T_0(t)$.
Since the study of Ciotti et al. (1991) does not include the mass
deposition due to local thermal instabilities in the ISM, one can
speculate, that their inherently unstable (either to wind or to
inflow) outflow regime, probably, corresponds to our, in the same
sense unstable, equilibrium recycling model.  In our case the hot ISM
in elliptical galaxies could be accumulating to fill the potential
well and to approach `thermodynamic' equilibrium between sinks and
sources, while the SN activity and stellar mass loss decrease.  The
evolution in the outflow phase near equilibrium could last for a long
time before it enters the inflow phase and ends up with a cooling
catastrophe.  It is expected that this transition can also be
triggered earlier by external or internal processes which disturb the
ISM of the galaxy.  However, a thorough study of the stability for
evolutionary ISM models with time-dependent sources is needed in order
to draw rigorous conclusions for this non-linear problem.

Our hybrid inflow-outflow solutions indicate that the central cooling
catastrophe can develop in the form of radially oriented streams of
low density peripheral gas penetrating the subsonic outflow. The
vertical orientation of (nearly) isolated giant ellipticals is defined
by the condition $t_{\rmn{G}} \ll t_{\rmn{c,0}}$.  When the thermal
and gravitational time scales are comparable in the central region, or
when $t_{\rmn{G}}>t_{\rmn{c,0}}$, the condensation fronts will be
orientated in a random way, since gravity does not focus the flows.
This case is described in detail in \cite{kritsuk96a}.  The matter,
deposited in narrow streams, can show up in optical and UV emission
lines as filamentary nebulae residing within a radius of a few kpc
from the nucleus.  The morphology of such filaments depends both on
the ratio of the characteristic thermal and free fall time scales and
on the topological properties of the large-scale perturbations
triggering the instability.  A detailed study of the dynamics of the
cooling catastrophe requires multidimensional high resolution
simulations, which would allow one to follow the development of the
hydrodynamic instabilities further into the non-equilibrium epoch.
This work is in progress.

Numerical experiments allow us to identify two modes for mass
deposition in hot galactic coronae.  The first, hydrostatic continuous
mode, corresponds to a distributed mass deposition, as it occurs in
equilibrium models and in slow or subsonic outflows.  The second,
dynamical mode, is related to the catastrophic cooling regime.  These
modes can be responsible for the diffuse blue light emission due to
continuously distributed star formation of low efficiency and due to
massive young star clusters formed in bursts triggered by large-scale
perturbations.  The amount of mass deposited in both ways can be
comparable. However, the mean $\dot\rho_{\rmn{ti}}(r)$ distribution
follows the density of the old stellar population in the quiescent
mode, while it is much more concentrated towards the centre in case of
the catastrophic mode.

Currently our study does not include the effects of an additional
intra-cluster gas component, which typically has a considerably higher
temperature corresponding to the cluster gravitational potential.  The
intra-cluster medium can serve as a thermal energy reservoir for the
ISM of the galaxy and it can essentially modify the gas dynamics
through the enhanced pressure at the outer boundary.  We will address
these problems in a subsequent publication.  On the other hand, slow
rotation and flattening of the galactic gravitational potential do
introduce finite perturbations to the equilibrium model.  Hence, one
expects a hybrid solution with inflow along the equatorial plane and
with polar outflows.  Recent results of D'Ercole \& Ciotti (1997)
\nocite{dercole.97} have clearly demonstrated that this is indeed the
case.

\section{Conclusions and perspective}
Using numerical technique we probed the stability of the hydrostatic
equilibrium recycling model for hot gaseous coronae of giant
elliptical galaxies with respect to a variety of perturbations.

In the absence of heat conduction the equilibrium appears to be
unstable on a time scale of the order of the thermal time for the gas
at the galactic centre.  The physical reason for this instability is
rooted in the properties of the source terms.  The energy and mass
sinks due to local thermal instabilities essentially modify the
elastic properties of the gas.  A slow contraction of a gas element
amplifies the condensation and cooling in it and, thereby, the mass
inflow into the compressed region.  As a result, the density rises,
providing further growth of losses and accelerating the contraction in
a runaway regime.  A slow expansion of the gas element instead
decreases the losses and increases the specific heating per unit mass,
which further decreases the losses and accelerates the element
expansion. The expansion and heating saturate as the gas temperature
approaces the characteristic temperature of the heat source. 
Both types of unstable behaviour can be identified in our
simulations.

Spherically symmetric perturbations trigger either gas inflow, which
ends up with a cooling catastrophe in the core of a galaxy, or a
subsonic outflow regime, which efficiently removes the gas from the
central region.  In the first case the X-ray surface brightness at the
centre of the galaxy considerably exceeds observed values, while in
the second case the hot gas is practically invisible in X-rays.

Thermal conduction is able to maintain a stable equilibrium only
against low-amplitude perturbations.  Spherically symmetric, global
density perturbations with amplitudes $\ge10$~\% remain unstable.  In
case of inflows the conductive heat flux cannot prevent the cooling
catastrophe at the centre, but can delay it in time considerably.  On
the other hand, when the thermal conductivity is not suppressed, the
gas velocities of outflows saturate at much lower values. This gives
rise to quasi-steady-state solutions with an X-ray luminosity only
slightly lower than that of the equilibrium model.

Using axisymmetric perturbations, we were able to study the 2D
hydrodynamics of the cooling catastrophe for the first time.  In axial
symmetry the system exhibits an qualitatively different behaviour. A
set of narrow cooling gas streams, flowing towards the centre through
a global subsonic galactic wind, develops in response to random
velocity perturbations of the equilibrium recycling model.  In the
strongly non-linear regime the characteristic averaged X-ray
observables of such hybrid gas flows (e.g., the surface brightness and
the temperature profiles) mimic the characteristics of the initial
equilibrium state, while the equilibrium itself appears to be
considerably violated.

These results indicate the need for a three-dimensional time-dependent
treatment of the problem, which would be able to reveal non-linear
instability saturation mechanisms.  At the same time, they cast doubts
on using steady-state `cooling flow' type models, based on the
assumption of quasi-hydrostatic equilibrium, for recovering mass
deposition rates in the central galactic regions.

\section*{Acknowledgments}
This work was partly supported by the Russian Foundation for Basic
Research (project 96-02-19670) and by the Federal Targeted Programme
{\em Integration} (project 578).  A.K. is grateful to Eugene Churazov
for stimulating discussions, and to the staff of the MPA and MPE for
their warm hospitality.

%\bibliography{../Bibliography/cflow,../Bibliography/ellip,../Bibliography/hydro}
%\bibliographystyle{../Styles/avtorgod}

\bsp
\label{lastpage}

\end{document}